\documentclass[12pt]{article}
\usepackage[utf8]{inputenc}
\usepackage{times}
\usepackage{bm} 

\title{Pump-shaping of non-collinear and non-degenerate entangled photons}
\author{Ohad Lib, Yaron Bromberg$^*$}

\date{}

\usepackage{graphicx}
\usepackage{float} 
\usepackage[sorting = none, backend = bibtex, style=numeric-comp]{biblatex}
\begin{document}

\maketitle
\noindent{Racah Institute of Physics, The Hebrew University of Jerusalem, Jerusalem, 91904 Israel}

\noindent{*Yaron.Bromberg@mail.huji.ac.il}
\begin{abstract}
Free-space quantum key distribution is gaining increasing interest as a leading platform for long range quantum communication. However, the sensitivity of quantum correlations to scattering induced by turbulent atmospheric links limits the performance of such systems. Recently, methods for compensating the scattering of entangled photons have been demonstrated, allowing for real-time optimization of their quantum correlations. In this Letter, we demonstrate the use of wavefront shaping for compensating the scattering of non-collinear and non-degenerate entangled photons. These results demonstrate the applicability of wavefront shaping schemes for protocols utilizing the large bandwidth and emission angle of the entangled photons.
\end{abstract}
In recent years, quantum technologies are starting to make the transition to commercial, real-world, applications. One of the most promising applications is quantum key distribution (QKD), where photonic quantum bits (qubits) are typically used for secure sharing of a secret key between two parties\cite{bennett1984quantum}. The key distribution is usually performed using single mode fiber links\cite{chen2020sending}, or more recently, free-space links, for example between a ground station and a satellite\cite{liao2017satellite,yin2017satellite}. Free-space QKD is in particular attractive, as free-space links exhibit lower loss than optical fibers, enabling communication over large distances. Nevertheless, implementation of free-space links in real-world scenarios is challenging, especially in the presence of turbulence in the atmosphere, which results in the scattering of the photons\cite{liao2017satellite,ursin2007entanglement}. 

A promising approach to compensate the scattering of single and entangled photons is to borrow concepts from the well-established fields of adaptive optics\cite{tyson2015principles} and wavefront shaping\cite{vellekoop2007focusing,mosk2012controlling}, and apply them to quantum links. To this end, methods for shaping the wavefront of single and entangled photons using spatial light modulators (SLMs) have been demonstrated, either by controlling the photons directly\cite{defienne2016two,wolterink2016programmable,defienne2018adaptive}, or tailoring the classical pump beam that stimulates their creation via spontaneous parametric down conversion (SPDC)\cite{minozzi2013optimization,pugh2016towards,lib2020real}. Recently, a pump-shaping method in which the intensity of the classical pump beam is used as feedback for compensating the scattering of entangled photons has been demonstrated, allowing for real-time optimization of entangled photons\cite{lib2020real}. However, this demonstration was realized only in a collinear and frequency-degenerate configuration of SPDC, which limits the range of possible applications. Generalizing pump-shaping to non-degenerate and non-collinear configurations, requires special care, as non-degenerate photons accumulate different phases in the random medium, and non-collinear photons do not propagate in the same direction as the pump beam. 

In this Letter, we demonstrate the use of pump-shaping in non-collinear and non-degenerate configurations of SPDC. For non-collinear SPDC, we show that the intensity of the pump beam can still be used as feedback for optimizing the quantum correlations, as long as the emission angle of the SPDC light is within the memory effect, or equivalently, isoplanatic patch of the scattering medium\cite{freund1988memory,beckers1988increasing}. We experimentally demonstrate this result by optimizing the correlations between entangled photons generated via non-collinear SPDC and scattered by a thin diffuser. For non-degenerate SPDC, we show that even though the wavelengths of the two entangled photons are different, energy conservation ensures the surprising correspondence between the intensity of the pump beam and the quantum correlations between the entangled photons, allowing the use of the pump-shaping method\cite{lib2020real}. We experimentally demonstrate optimization of entangled photons separated by nearly $100 nm$. We believe these results will increase the range of applications that can benefit from quantum wavefront shaping, for example, in protocols utilizing the large number of spectral\cite{pseiner2020experimental} and spatial modes\cite{walborn2006quantum} of SPDC for multiplexing quantum information.

We begin by establishing the correspondence between the scattering of the classical pump beam and the entangled photons, coined \textit{signal} and \textit{idler} photons. We generalize the derivation in Ref.\cite{lib2020real} for the cases of non-collinear and non-degenerate entangled photons. Generally, the quantum state of the entangled photons is given by $\\|{\psi}\rangle=\int d \mathbf{q_s}d\mathbf{q_i}d\omega_sd\omega_i\psi(\mathbf{q_s,q_i},\omega_s,\omega_i)a^\dagger(\mathbf{q_s},\omega_s)a^\dagger(\mathbf{q_i},\omega_i)|0\rangle$ where $a^\dagger(\mathbf{q},\omega)$ is the creation operator of a photon with a transverse wavevector $\mathbf{q}$ and frequency $\omega$, $|0\rangle$ is the vacuum state, and we assume the signal and idler photons have the same polarization\cite{walborn2010spatial}. For a thin crystal pumped with a monochromatic beam with frequency $\omega_p$, one can neglect the effect of phase matching and write the two photon amplitude as\cite{walborn2010spatial,monken1998transfer}
\begin{equation}\label{1}
\psi(\mathbf{q_s,q_i},\omega_s,\omega_i) \propto \delta(\omega_s+\omega_i-\omega_p)v\left(\mathbf{q_s+q_i}\right),
\end{equation}
where $v(\mathbf{q})$ is the angular spectrum of the pump beam. We model a thin diffuser placed at the plane of the nonlinear crystal as an amplitude transfer function of the form $A_d(\bm{\rho},\omega)=exp(ih(\bm{\rho})(n-1)\omega/c)$, where $\bm{\rho}$ is the transverse spatial coordinate, $n$ is the refractive index of the diffuser which is assumed to be weakly dependent on frequency, $c$ is the speed of light in vacuum and $h(\bm{\rho})$ is the position dependent thickness of the diffuser. The rate of coincidence events between photons with transverse wavevectors $\mathbf{q_{s,i}}$ and frequencies $\omega_{s,i}$ is then given by\cite{lib2020real}

\begin{equation}\label{2}
C(\mathbf{q_s,q_i},\omega_s,\omega_p-\omega_s)\propto \left|\int d\bm{\rho}W(\bm{\rho})A_d(\bm{\rho},\omega_s)A_d(\bm{\rho},\omega_p-\omega_s)\exp(-i\bm{\rho}\cdot(\mathbf{q_s+q_i}))\right|^2
\end{equation}

where $W(\bm{\rho})$ is the profile of the pump beam at the nonlinear crystal, and $\omega_i=\omega_p-\omega_s$ must hold for any non-zero coincidence rate. Interestingly, as $A_d(\bm{\rho},\omega_p)=A_d(\bm{\rho},\omega_s)A_d(\bm{\rho},\omega_p-\omega_s)$, the coincidence rate corresponds to the intensity of the pump beam in the direction $\mathbf{q}=\mathbf{q_i}+\mathbf{q_s}$. As the correspondence between the quantum correlations and the intensity of the pump beam holds for any pump profile $W(\bm{\rho})$ (Eq.\ref{2}), by optimizing the shape of the classical pump beam, the scattering of the entangled photons is simultaneously compensated, regardless of the frequency separation between them. It is worth noting that pump-shaping techniques were also recently employed for controlling the output beams in second harmonic generation\cite{libster2015dynamic,thompson2017enhanced,liu2017dynamic,liu2018dynamic}.

For non-collinear configurations, one can simply consider the case of non-zero $\mathbf{q_i}$ in Eq.\ref{2}. For a thin diffuser, the shape of the output speckle pattern is independent of the incident angle, therefore by focusing the pump beam at $\mathbf{q}=0$, the coincidence rate will be localized at $\mathbf{q_s}=-\mathbf{q_i}$. For thick scattering media, the scattering properties are in general angle dependent. However, there is always a finite range of angles for which the shape of the output speckle pattern remains unchanged, called the memory range or isoplanatic patch\cite{freund1988memory,beckers1988increasing}. Therefore, for a general diffuser, the optimization of the pump beam will simultaneously optimize the quantum correlations between the entangled photons, as long as their emission is within the angular range dictated by the memory effect or isoplanatic patch.

\begin{figure}[H]
\centering
\includegraphics[width=0.75\textwidth]{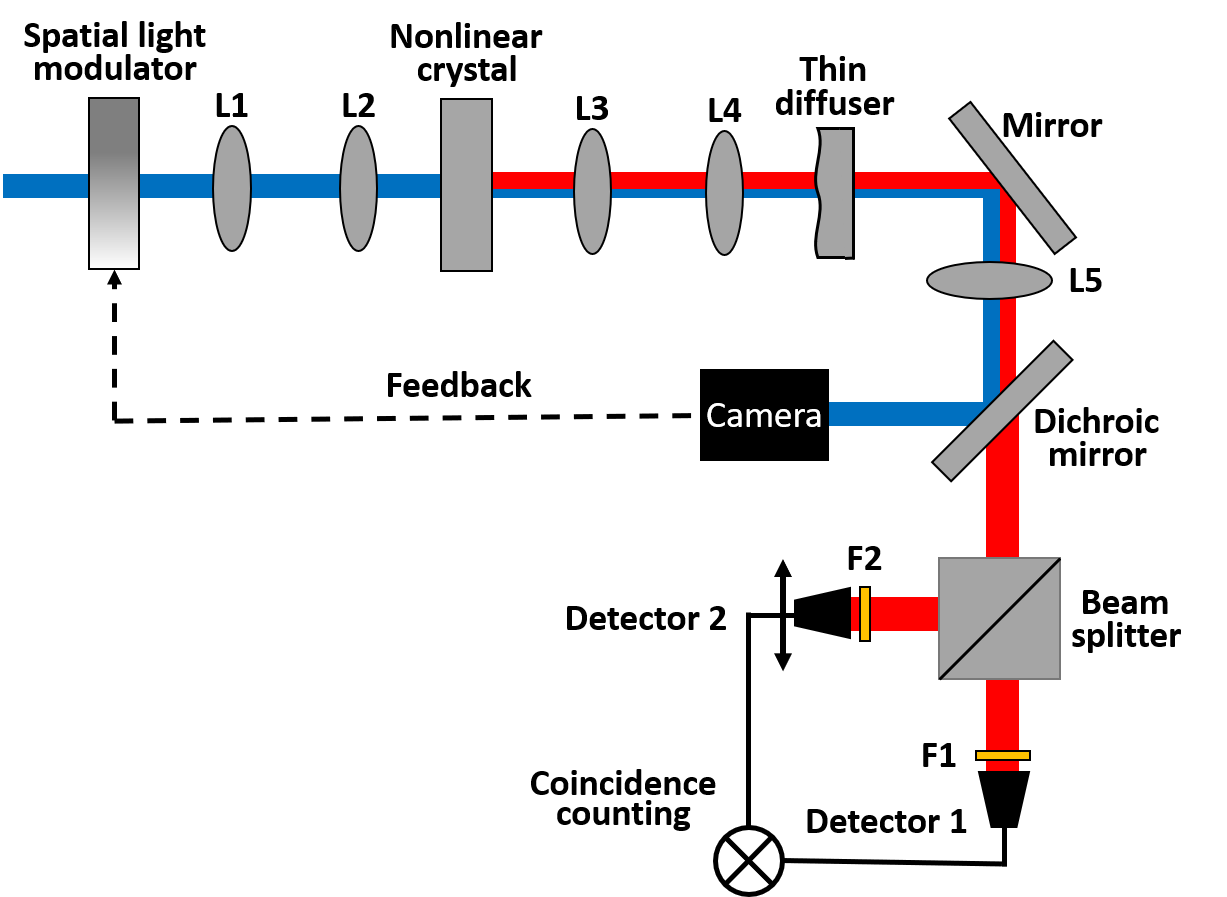}
  \caption{\label{fig:1}  \textbf{Experimental setup.} Entangled photons are generated via spontaneous parametric down conversion by pumping an PPKTP nonlinear crystal with a $404 nm$ pump beam. Both the pump beam and the entangled photons are imaged onto a thin diffuser. Interference filters are placed before the single photon detectors ($F1,F2$). Either two bandpass filters centered at $\lambda_i=\lambda_s=808 nm$ or $\lambda_i=766 nm$ and $\lambda_s=850nm$ are used for the degenerate and non-degenerate cases respectively. The full width at half maximum of the filters is approximately $10 nm$. A phase-only spatial light modulator is used for tailoring the wavefront of the pump beam to compensate the scattering induced by the thin diffuser.}
\end{figure}

The experimental setup is presented in fig.\ref{fig:1}. A continuous-wave laser ($404 nm, 50 mW$) is shaped using a phase-only SLM. The shaped pump beam is imaged onto a $2 mm$ long nonlinear PPKTP crystal, generating spatially entangled photons via SPDC. Both the entangled photons and the classical pump beam are then imaged onto a thin polymer-on-glass diffuser (divergence angle $0.25^o$). The pump beam and the entangled photons are separated using a dichroic mirror, and measured at the far-field using an CMOS camera and filtered single-photon detectors, respectively. Coincidence measurements are performed by scanning the position of the signal detector, while keeping the idler detector stationary.

\begin{figure}[h!]
\centering
\includegraphics[width=0.25\textwidth]{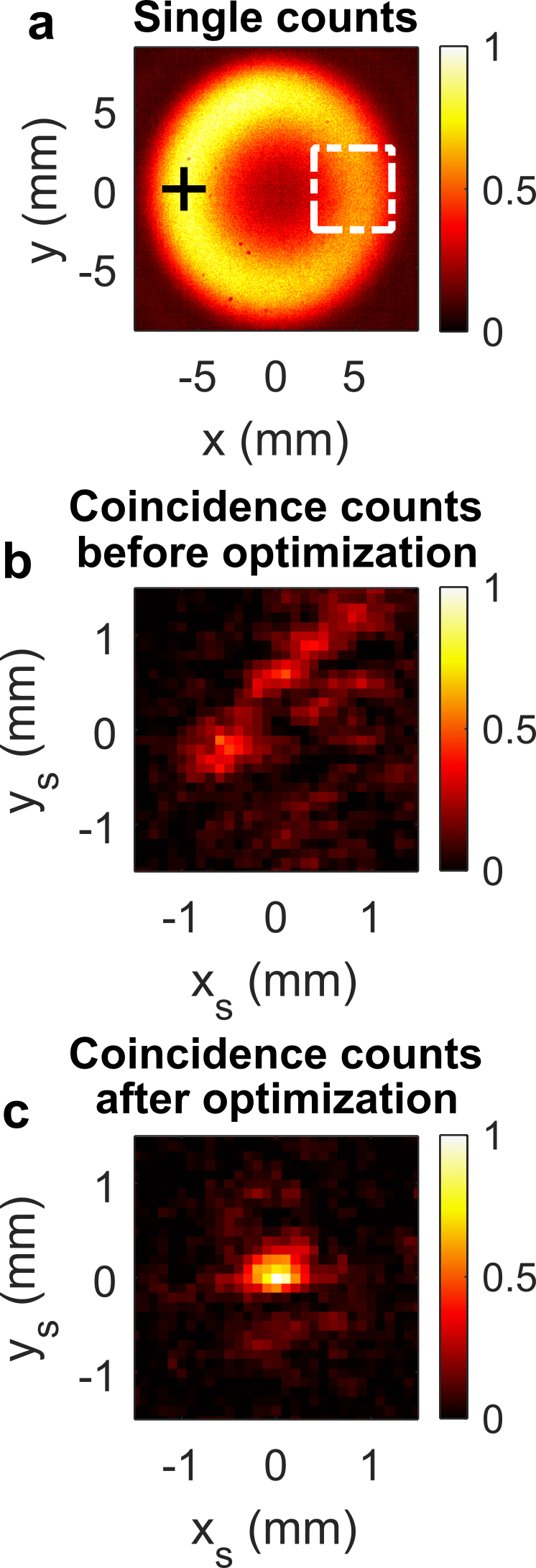}
  \caption{\label{fig:2}  \textbf{Non-collinear SPDC.} (a) Far-field single photon counts with non-collinear phase matching. The positions of the stationary idler detector and the scanning area of the signal detector are marked with a black cross and a white square, respectively. (b) Without optimization, the coincidence rate features a two-photon speckle pattern. (c) By optimizing the intensity of the pump beam (not shown), the quantum correlations are simultaneously localized.}
\end{figure}

First, we consider the case of non-collinear SPDC. The temperature of the PPKTP crystal is set such that non-collinear phase matching is obtained, which is manifested in a ring-shaped single photon counts distribution at the far-field (fig.\ref{fig:2}a). Before the optimization, the coincidence pattern exhibits a two-photon speckle pattern due to scattering induced by the diffuser\cite{peeters2010observation} (fig.\ref{fig:2}b), where the signal detector scans positions (white square) around the symmetric point to the stationary idler detector (black cross). Then, using the SLM and the partitioning optimizing algorithm\cite{vellekoop2008phase}, the pump beam is re-focused at the far-field, and the quantum correlations are simultaneously enhanced (fig.\ref{fig:2}c). In this measurement, $10 nm$ wide interference filters centered around $808 nm$ are used. This result demonstrates the applicability of the pump optimization approach for quantum wavefront shaping with non-collinear SPDC.

\begin{figure}[h!]
\centering
\includegraphics[width=0.7\textwidth]{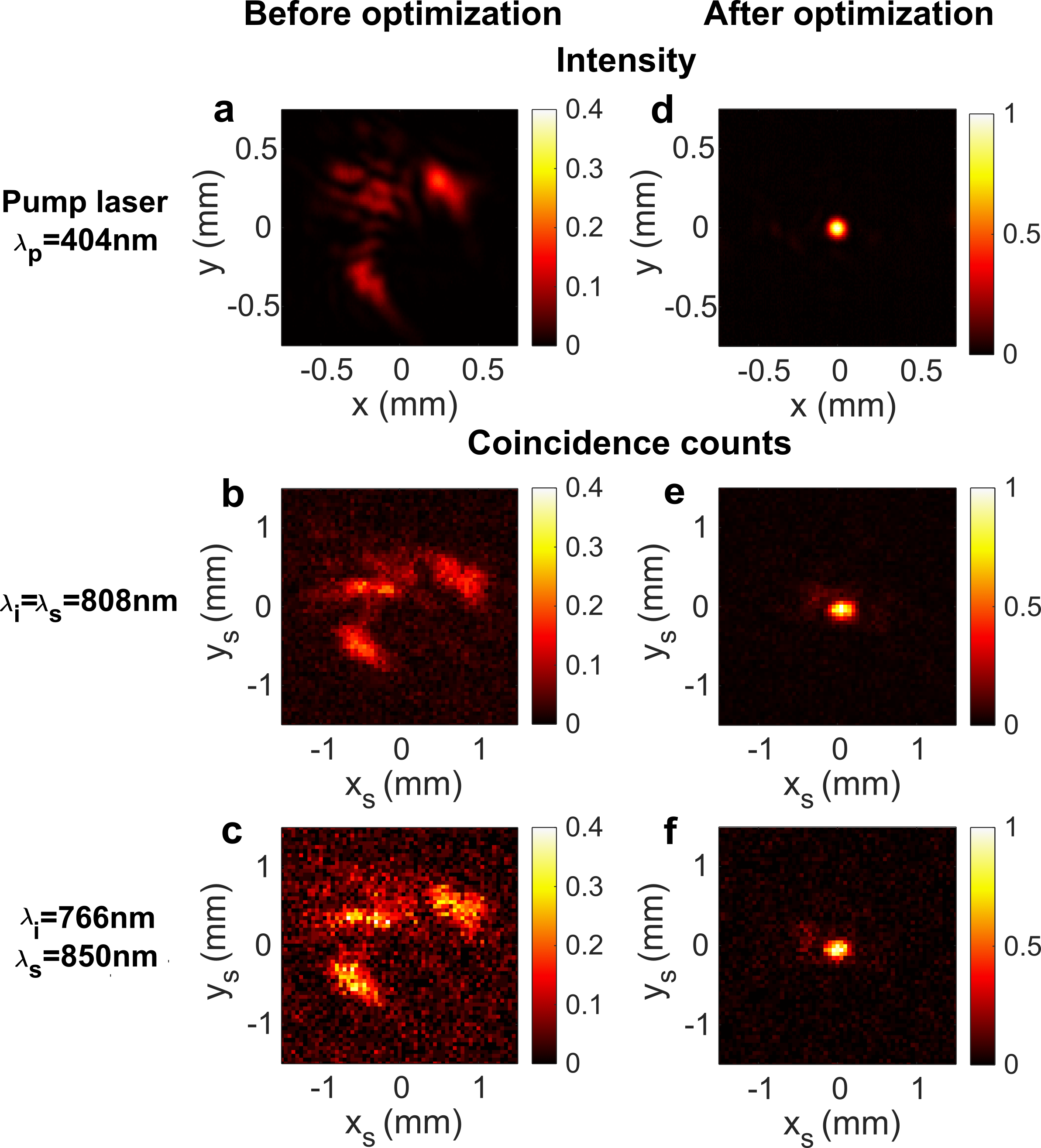}
  \caption{\label{fig:3}  \textbf{Non-degenerate SPDC.} Similar speckle patterns are observed for the intensity of the pump beam ($\lambda_p=404nm$) (a) and the quantum correlations between the degenerate ($\lambda_i=\lambda_s=808nm$) (b) and non-degenerate ($\lambda_i=766nm, \lambda_s=850nm$) (c) entangled photons. By optimizing the intensity of the pump beam using classical wavefront shaping (d), the scattering of the degenerate (e) and non-degenerate (f) entangled photons is compensated as well. }
\end{figure}

Next, we consider the case of non-degenerate SPDC. In this experiment, the temperature of the crystal is set for a collinear phase matching, yet a non-degenerate configuration is measured as well, where the idler photon is filtered using a $13 nm$ wide filter centered at $766 nm$, and the signal photon using a $10 nm$ wide filter centered at $850 nm$. A clear similarity between the speckle pattern of the pump beam (fig.\ref{fig:3}a) and the two-photon speckle patterns is observed in both the degenerate (fig.\ref{fig:3}b) and non-degenerate (fig.\ref{fig:3}c) configurations, in agreement with Eq.\ref{2}. The lower signal to noise ratio in the non-degenerate case is a result of the lower signals associated with the measured wavelengths, together with a reduction by a factor of two caused by the use of a simple beam splitter rather than a dichroic mirror to separate the non-degenerate entangled photons. By compensating the scattering of the classical pump beam (fig.\ref{fig:3}d), the quantum correlations between both the frequency degenerate (fig.\ref{fig:3}e) and non-degenerate (fig.\ref{fig:3}f) entangled photons are optimized.

It is interesting to note that while all three speckle patterns are similar in shape, they exhibit different scales due to their different associated wavelengths. The two-photon speckle pattern of the degenerate $808 nm$ photons is twice as large compared with that of the $404 nm$ pump beam\cite{lib2020real}. For the non-degenerate case, the question of scale is more interesting, as in principal one has both the wavelengths of the signal and idler photons to consider. However, although this effect is too subtle to be appreciated in our measurements, by looking at Eq.\ref{2} one can see that the scale of the coincidence pattern is only determined by the wavelength measured by the scanning detector, which is $850 nm$ in our case. Thus, in contrast to the degenerate case, the scale on the coincidence pattern is dependent on which of the two detectors is scanning and which is stationary.

In conclusion, we have demonstrated the applicability of quantum wavefront shaping with the recently proposed pump-based approach, for non-collinear and non-degenerate entangled photons. We have shown that for non-collinear configurations, the feedback provided by the classical pump beam can still be used, given that the emission angle of the SPDC light is within the range allowed by the memory effect or the isoplanatic patch. For entangled photons having non-degenerate frequencies, we have shown that due to energy conservation in SPDC, the intensity of the pump beam can be used as feedback for optimizing the quantum correlations in this case as well. We experimentally demonstrated the optimization of non-degenerate entangled photons separated by nearly $100 nm$. These results pave the way towards implementation of wavefront shaping techniques in a wide variety of configurations. In particular, non-collinear SPDC configurations can be used for various satellite-based QKD schemes, where the addition of quantum wavefront shaping technology could help mitigate the effect of scattering induced by the atmosphere\cite{aspelmeyer2003long,bedington2017progress}. In addition, we believe that the ability to compensate the scattering of large-bandwidth SPDC could be important for recent protocols relying on spectral multiplexing for increasing the capacity of QKD links\cite{pseiner2020experimental}.

\noindent\textbf{Funding.} Zuckerman STEM Leadership Program; Israel Science Foundation  (1268/16); United States-Israel Binational Science Foundation (BSF) (2017694).

\printbibliography

@article{bennett1984quantum,
  title={Quantum cryptography: Public key distribution and coin tossing},
  author={Bennett, Charles H and Brassard, Gilles},
  journal={Proc. Int. Conf. on Computers, Systems and Signal Processing},
  year={1984}
}

@article{chen2020sending,
  title={Sending-or-Not-Sending with Independent Lasers: Secure Twin-Field Quantum Key Distribution over 509 km},
  author={Chen, Jiu-Peng and Zhang, Chi and Liu, Yang and Jiang, Cong and Zhang, Weijun and Hu, Xiao-Long and Guan, Jian-Yu and Yu, Zong-Wen and Xu, Hai and Lin, Jin and others},
  journal={Physical review letters},
  volume={124},
  number={7},
  pages={070501},
  year={2020},
  publisher={APS}
}

@article{liao2017satellite,
  title={Satellite-to-ground quantum key distribution},
  author={Liao, Sheng-Kai and Cai, Wen-Qi and Liu, Wei-Yue and Zhang, Liang and Li, Yang and Ren, Ji-Gang and Yin, Juan and Shen, Qi and Cao, Yuan and Li, Zheng-Ping and others},
  journal={Nature},
  volume={549},
  number={7670},
  pages={43--47},
  year={2017},
  publisher={Nature Publishing Group}
}

@article{yin2017satellite,
  title={Satellite-based entanglement distribution over 1200 kilometers},
  author={Yin, Juan and Cao, Yuan and Li, Yu-Huai and Liao, Sheng-Kai and Zhang, Liang and Ren, Ji-Gang and Cai, Wen-Qi and Liu, Wei-Yue and Li, Bo and Dai, Hui and others},
  journal={Science},
  volume={356},
  number={6343},
  pages={1140--1144},
  year={2017},
  publisher={American Association for the Advancement of Science}
}

@article{ursin2007entanglement,
  title={Entanglement-based quantum communication over 144 km},
  author={Ursin, Rupert and Tiefenbacher, F and Schmitt-Manderbach, T and Weier, H and Scheidl, Thomas and Lindenthal, M and Blauensteiner, B and Jennewein, T and Perdigues, J and Trojek, P and others},
  journal={Nature physics},
  volume={3},
  number={7},
  pages={481--486},
  year={2007},
  publisher={Nature Publishing Group}
}

@article{mosk2012controlling,
  title={Controlling waves in space and time for imaging and focusing in complex media},
  author={Mosk, Allard P and Lagendijk, Ad and Lerosey, Geoffroy and Fink, Mathias},
  journal={Nature photonics},
  volume={6},
  number={5},
  pages={283--292},
  year={2012},
  publisher={Nature Publishing Group}
}

@book{tyson2015principles,
  title={Principles of adaptive optics},
  author={Tyson, Robert K},
  year={2015},
  publisher={CRC press}
}

@article{vellekoop2007focusing,
  title={Focusing coherent light through opaque strongly scattering media},
  author={Vellekoop, Ivo M and Mosk, AP},
  journal={Optics letters},
  volume={32},
  number={16},
  pages={2309--2311},
  year={2007},
  publisher={Optical Society of America}
}

@article{defienne2016two,
  title={Two-photon quantum walk in a multimode fiber},
  author={Defienne, Hugo and Barbieri, Marco and Walmsley, Ian A and Smith, Brian J and Gigan, Sylvain},
  journal={Science advances},
  volume={2},
  number={1},
  pages={e1501054},
  year={2016},
  publisher={American Association for the Advancement of Science}
}

@article{wolterink2016programmable,
  title={Programmable two-photon quantum interference in 10 3 channels in opaque scattering media},
  author={Wolterink, Tom AW and Uppu, Ravitej and Ctistis, Georgios and Vos, Willem L and Boller, Klaus-J and Pinkse, Pepijn WH},
  journal={Physical Review A},
  volume={93},
  number={5},
  pages={053817},
  year={2016},
  publisher={APS}
}

@article{defienne2018adaptive,
  title={Adaptive quantum optics with spatially entangled photon pairs},
  author={Defienne, Hugo and Reichert, Matthew and Fleischer, Jason W},
  journal={Physical review letters},
  volume={121},
  number={23},
  pages={233601},
  year={2018},
  publisher={APS}
}

@article{pugh2016towards,
  title={Towards correcting atmospheric beam wander via pump beam control in a down conversion process},
  author={Pugh, Christopher J and Kolenderski, Piotr and Scarcella, Carmelo and Tosi, Alberto and Jennewein, Thomas},
  journal={Optics Express},
  volume={24},
  number={18},
  pages={20947--20955},
  year={2016},
  publisher={Optical Society of America}
}

@article{lib2020real,
  title={Real-time shaping of entangled photons by classical control and feedback},
  author={Lib, Ohad and Hasson, Giora and Bromberg, Yaron},
  journal={Science Advances},
  volume={6},
  number={37},
  pages={eabb6298},
  year={2020},
  publisher={American Association for the Advancement of Science}
}

@article{freund1988memory,
  title={Memory effects in propagation of optical waves through disordered media},
  author={Freund, Isaac and Rosenbluh, Michael and Feng, Shechao},
  journal={Physical review letters},
  volume={61},
  number={20},
  pages={2328},
  year={1988},
  publisher={APS}
}

@inproceedings{beckers1988increasing,
  title={Increasing the size of the isoplanatic patch with multiconjugate adaptive optics},
  author={Beckers, Jacques M},
  booktitle={European Southern Observatory Conference and Workshop Proceedings},
  volume={30},
  pages={693},
  year={1988}
}

@article{walborn2006quantum,
  title={Quantum key distribution with higher-order alphabets using spatially encoded qudits},
  author={Walborn, SP and Lemelle, DS and Almeida, MP and Ribeiro, PH Souto},
  journal={Physical review letters},
  volume={96},
  number={9},
  pages={090501},
  year={2006},
  publisher={APS}
}

@article{pseiner2020experimental,
  title={Experimental wavelength-multiplexed entanglement-based quantum cryptography},
  author={Pseiner, Johannes and Achatz, Lukas and Bulla, Lukas and Bohmann, Martin and Ursin, Rupert},
  journal={arXiv preprint arXiv:2009.03691},
  year={2020}
}

@article{walborn2010spatial,
  title={Spatial correlations in parametric down-conversion},
  author={Walborn, Stephen P and Monken, CH and P{\'a}dua, S and Ribeiro, PH Souto},
  journal={Physics Reports},
  volume={495},
  number={4-5},
  pages={87--139},
  year={2010},
  publisher={Elsevier}
}

@article{monken1998transfer,
  title={Transfer of angular spectrum and image formation in spontaneous parametric down-conversion},
  author={Monken, Carlos Henrique and Ribeiro, PH Souto and P{\'a}dua, Sebasti{\~a}o},
  journal={Physical Review A},
  volume={57},
  number={4},
  pages={3123},
  year={1998},
  publisher={APS}
}

@article{peeters2010observation,
  title={Observation of two-photon speckle patterns},
  author={Peeters, WH and Moerman, JJD and Van Exter, MP},
  journal={Physical review letters},
  volume={104},
  number={17},
  pages={173601},
  year={2010},
  publisher={APS}
}

@article{vellekoop2008phase,
  title={Phase control algorithms for focusing light through turbid media},
  author={Vellekoop, Ivo Micha and Mosk, AP},
  journal={Optics communications},
  volume={281},
  number={11},
  pages={3071--3080},
  year={2008},
  publisher={Elsevier}
}

@article{aspelmeyer2003long,
  title={Long-distance quantum communication with entangled photons using satellites},
  author={Aspelmeyer, Markus and Jennewein, Thomas and Pfennigbauer, Martin and Leeb, Walter R and Zeilinger, Anton},
  journal={IEEE Journal of Selected Topics in Quantum Electronics},
  volume={9},
  number={6},
  pages={1541--1551},
  year={2003},
  publisher={IEEE}
}

@article{bedington2017progress,
  title={Progress in satellite quantum key distribution},
  author={Bedington, Robert and Arrazola, Juan Miguel and Ling, Alexander},
  journal={npj Quantum Information},
  volume={3},
  number={1},
  pages={1--13},
  year={2017},
  publisher={Nature Publishing Group}
}

@article{minozzi2013optimization,
  title={Optimization of two-photon wave function in parametric down conversion by adaptive optics control of the pump radiation},
  author={Minozzi, M and Bonora, S and Sergienko, AV and Vallone, G and Villoresi, P},
  journal={Optics letters},
  volume={38},
  number={4},
  pages={489--491},
  year={2013},
  publisher={Optical Society of America}
}

@article{libster2015dynamic,
  title={Dynamic control of light beams in second harmonic generation},
  author={Libster-Hershko, Ana and Trajtenberg-Mills, Sivan and Arie, Ady},
  journal={Optics Letters},
  volume={40},
  number={9},
  pages={1944--1947},
  year={2015},
  publisher={Optical Society of America}
}

@article{thompson2017enhanced,
  title={Enhanced second harmonic generation efficiency via wavefront shaping},
  author={Thompson, Jonathan V and Hokr, Brett H and Throckmorton, Graham A and Wang, Dawei and Scully, Marlan O and Yakovlev, Vladislav V},
  journal={ACS Photonics},
  volume={4},
  number={7},
  pages={1790--1796},
  year={2017},
  publisher={ACS Publications}
}

@article{liu2017dynamic,
  title={Dynamic computer-generated nonlinear-optical holograms},
  author={Liu, Haigang and Li, Jun and Fang, Xiangling and Zhao, Xiaohui and Zheng, Yuanlin and Chen, Xianfeng},
  journal={Physical Review A},
  volume={96},
  number={2},
  pages={023801},
  year={2017},
  publisher={APS}
}

@article{liu2018dynamic,
  title={Dynamic computer-generated nonlinear optical holograms in a non-collinear second-harmonic generation process},
  author={Liu, Haigang and Zhao, Xiaohui and Li, Hui and Zheng, Yuanlin and Chen, Xianfeng},
  journal={Optics Letters},
  volume={43},
  number={14},
  pages={3236--3239},
  year={2018},
  publisher={Optical Society of America}
}
\end{document}